\documentclass[9pt, technote]{IEEEtran}

\IEEEoverridecommandlockouts

\usepackage{cite}
\usepackage{amsmath,amssymb,amsfonts}
\usepackage{algorithmic}
\usepackage{algorithm}
\usepackage{graphicx}
\usepackage[dvipsnames]{xcolor}
\usepackage{booktabs} 
\usepackage[]{units} 
\usepackage{subcaption}
\usepackage[short,c2,nocomma]{optidef}
\usepackage{hyperref} 
\usepackage{tabularx}
\hypersetup{
    colorlinks,
    linkcolor={blue!60!black},
    citecolor={blue!60!black},
    urlcolor={black}
}
\definecolor{dgreen}{RGB}{50, 180, 0}
\title{
Mitigating Degradation Attacks in Cooperative Autonomous Driving via Intention Sharing: A Vehicle-in-the-Loop Study
}

\author{Prakhar Gupta$^{1}$, Tyler Ard$^{2}$, Rongyao Wang$^{1}$, Jagruti Sahoo$^{3}$, Judith Mwakalonge$^{4}$, Ardalan Vahidi$^{5}$, Yunyi Jia$^{1}$
\thanks{Copyright (c) 2026 IEEE. Personal use of this material is permitted. However, permission to use this material for any other purposes must be obtained from the IEEE by sending a request to pubs-permissions@ieee.org}
\thanks{This work is based upon the work supported by the National Center for Transportation Cybersecurity and Resiliency (TraCR) (a U.S. Department of Transportation National University Transportation Center) headquartered at Clemson University, Clemson, South Carolina, USA. Any opinions, findings, conclusions, and recommendations expressed in this material are those of the author(s) and do not necessarily reflect the views of TraCR, and the U.S. Government assumes no liability for the contents or use thereof.}
\thanks{Manuscript received 22 Apr 2026, Revised 15 Aug 2026, Accepted 8 Sep 2026.}
\thanks{$^{1}$Prakhar Gupta ({\tt\footnotesize prakhag@clemson.edu}), Rongyao Wang ({\tt\footnotesize rongyao@clemson.edu}), and Yunyi Jia ({\tt\footnotesize yunyij@clemson.edu}) are with the Dept. of Automotive Engineering, Clemson University, Greenville, SC 29607, USA.}%
\thanks{$^{2}$Tyler Ard ({\tt\footnotesize tard@anl.gov}) is with the Vehicle and Mobility Systems Department, Argonne National Laboratory, Lemont, IL 64039, USA.}%
\thanks{$^{3}$Jagruti Sahoo ({\tt\footnotesize jsahoo@scsu.edu}) is with the Computer Science \& Mathematics Department, South Carolina State University, Orangeburg, SC, USA.}
\thanks{$^{4}$Judith Mwakalonge ({\tt\footnotesize jmwakalo@scsu.edu}) is with the Engineering Department, South Carolina State University, Orangeburg, SC, USA.}%
\thanks{$^{5}$Ardalan Vahidi ({\tt\footnotesize avahidi@clemson.edu}) is with the Dept. of Mechanical Engineering, Clemson University, Clemson, SC 29634, USA.}%
}

\begin{document}
\bstctlcite{IEEEexample:BSTcontrol}

\maketitle
\thispagestyle{empty}
\pagestyle{empty}

\begin{abstract}
Communication delays induced by cyber attacks present a critical challenge to the safe operation of connected autonomous driving. This study investigates the use of intention sharing communication strategy to enhance the resilience of model predictive controllers under Denial-of-Service attacks. 
We employ a vehicle-in-the-loop testbed integrating a real drive-by-wire vehicle with a microscopic traffic simulator and vehicle-to-X communication infrastructure. We emulate Denial-of-service attacks that induce communication delays of up to two seconds.
We evaluate three controller variants: baseline status-sharing, intention-sharing, and delay-aware intention-sharing control.
Experimental results reveal that while baseline control suffers significant performance degradation and frequent collisions under adversarial delay, intention sharing eliminates collisions and maintains behavior near nominal levels for the tested scenarios. These findings demonstrate the practical potential of intention-sharing architectures for safeguarding connected vehicles against network-layer degradation.
\end{abstract}


\section{Introduction} \label{sec:intro}
Connected autonomous vehicles (CAVs) enhance road safety and efficiency through coordination strategies like cooperative car following. By leveraging vehicle-to-vehicle (V2V) communication to share state information, these systems outperform traditional adaptive cruise control in safety and energy efficiency \cite{ard2021energy, VAHIDI2018822, milanes2014cooperative, turri2016cooperative}.

However, wireless reliance introduces vulnerabilities to network degradation attacks. Among these, Denial-of-Service (DoS) attacks are especially concerning \cite{trkulja2020denialofserviceattackscv2xnetworks} and are among the most common cyber-attacks.
In such attacks, adversaries flood communication links with illegitimate traffic, leading to CPU and memory exhaustion at network nodes, thereby delaying or stalling legitimate V2X communication. Unlike jamming, which immediately blocks communication, these attacks can subtly degrade service quality by delaying or stalling the V2X communications. Recent transportation research demonstrates that even nominal delays can impact the behavior of cooperative driving and increase safety risks
\cite{behal2017characterisation, Khattak2023, jia2016platoon, PETHO2022100514}. Under adversarial conditions, stale information can cause control-model mismatch, erratic gap regulation, and rear-end collision risks, as also indicated by studies of connected-vehicle systems with communication delays \cite{petit2015potential,GE201446,Orosz02082016}.
These challenges can be addressed from two complementary perspectives: cyber-security mechanisms for attack detection, prevention, and mitigation; and control and communication architectures that preserve acceptable closed-loop behavior under the resulting network degradation. This study focuses on the latter perspective.

Among the control architectures that have been proposed, the Model Predictive Control (MPC) framework is especially appealing, as it can explicitly handle constraints while optimizing trajectories. 
Resilient controllers have been developed explicitly for DoS attacks, including distributed and event-triggered MPC formulations
\cite{chen_dmpc,zeng2023robust}. Variants of distributed and robust MPC have been shown to mitigate delay and packet loss effects in simulation \cite{10375252}. These studies primarily represent DoS attacks as periods of communication unavailability and validate their controllers through numerical simulations, without using intention sharing as the mitigation mechanism.
Another promising strategy is trajectory-level communication, such as in intention sharing. Here, each vehicle communicates its planned trajectory over a preview horizon rather than only current states \cite{intentOrosz, rong2024hybrid}. 
This can provide vehicles with predictive context and time-domain redundancy, potentially maintaining safety even when communication is delayed \cite{GUANETTI201818, guo2021anticipative}. 
Study in \cite{wang2023} investigated status and intent sharing under modeled V2X communication and vehicle-dynamics delays using numerical simulations driven by real highway data. In a separate study, \cite{intentOrosz} implemented intent messages using
commercial V2X radios and tested their packet-delivery performance
with real vehicles on public highways; however, the resulting
maneuver-level benefits were evaluated numerically rather than
through closed-loop vehicle experiments.
Lastly, promising experimental Vehicle-In-the-Loop (VIL) studies have quantified the effects of non-adversarial communication delay on CAV safety, stability, and energy performance \cite{Khattak2023,LI2024_jacky}, but do not evaluate intention sharing as a mitigation for adversarially induced delay.

Despite the theoretical benefits, experimental evidence quantifying the resilience of intention sharing under adversarial network degradation remains limited. Table~\ref{tab:novelty} compares the closest studies across these three dimensions of validation, communication degradation and intention sharing.
The distinguishing contribution of this work is therefore not any one of these elements in isolation, but their joint evaluation: intention sharing is investigated as a controller-level mitigation for experimental DoS-induced delays in a closed-loop framework incorporating a physical vehicle and V2X hardware.
\begin{table}[h]
\centering
\caption{Contrast with the closest studies relative to intention
sharing, communication degradation, and physical validation.}
\label{tab:novelty}
\setlength{\tabcolsep}{3.0pt}
\renewcommand{\arraystretch}{1.08}
\resizebox{\columnwidth}{!}{
\begin{tabular}{lccc}
\toprule
\textbf{Study} &
\textbf{Intent sharing} &
\textbf{Network condition} &
\textbf{Validation} \\
\midrule
Wang et al. \cite{wang2023}
    & Yes & Modeled V2X delays & Real data simulation \\

Wang et al. \cite{intentOrosz}
    & Yes & Non-adversarial packet loss & V2X + simulation \\

Ard et al. \cite{ard2021energy}
    & Yes & Nominal & VIL \\

Chen et al. \cite{chen_dmpc}
    & No & DoS communication loss & Simulation \\

Zeng et al. \cite{zeng2023robust}
    & No & DoS channel blocking & Simulation \\

Khattak et al. \cite{Khattak2023}
    & No & Modeled V2X delays & Field data \\

Li et al. \cite{LI2024_jacky}
    & No & Modeled V2X delays & VIL/field \\

\textbf{This work}
    & \textbf{Yes} & \textbf{DoS-induced delays}
    & \textbf{VIL + V2X} \\
\bottomrule
\end{tabular}}
\end{table}

Specifically, we investigate the impact of network-induced delay on cooperative car-following performance using an in-house developed VIL framework that integrates a microscopic traffic simulator, a real drive-by-wire vehicle, and a physical V2X communication stack. 
We emulate degradation using denial-of-service attacks, which induce queuing delays without total disconnection.
We evaluate three communication strategies for the control loop, and system resilience is evaluated through experimentally measured speed tracking errors, inter-vehicle distance gaps, and collision events under both nominal and adversarial conditions. 
While prior studies have focused separately on delay effects in simulation, and intention sharing in nominal conditions, this work delivers experimental data to show that intention sharing can provide a practical and effective resilience layer for  car-following under the considered adversarial network delays.

The key contributions of this work are summarized below:
\begin{itemize}
    \item \textbf{Intention sharing mitigation strategy}: We formulate and experimentally evaluate a timestamp-based alignment mechanism that compensates for adversarial delays by using the received trajectory information and vehicle intentions, improving temporal consistency in the control loop.
    \item \textbf{First closed-loop VIL comparison}: To our knowledge, this constitutes the first VIL experimental study that focuses on and directly contrasts communication strategies like intention sharing against status sharing for connected controls under adversarial delay. We go beyond simulation by using full V2X stack for experimental threat emulation. This links network-layer behavior to control-layer safety outcomes.
    \item \textbf{Experimental evidence of resilience}: For the car-following trials in this study, we provide empirical data that shows intention sharing improves resilience relative to status sharing when exposed to the considered adversarial network delay conditions.
\end{itemize}

\section{Framework} \label{sec:fw}
\subsection{Control Framework for Car Following}
The Predictive Car-Following (PCF) formulation is inspired by \cite{ard2021energy} and stated as in Eq.~\eqref{eq:mpc}. 
Here, $\mathcal R$ is the terminal state reference tracked by the model predictive controller, obtained from the infinite-horizon LQR solution for following the preceding vehicle (PV) with constant time headway $T$. $Q$ is the weighing matrix for the terminal cost. Parameters $\rho>0$ and $w\gg 0$ penalize the control input $u_i$ and slack variables $\epsilon_j$, respectively, and $i$ denotes the stage index in the $N$-step prediction horizon. 
\begin{mini}
    {u_i, \epsilon_j }{
     \left\Vert x_N - \mathcal R \right\Vert_Q^2}
     + \sum_{i=0}^{N-1} \rho u_i^2 + \sum_{j=0}^2 w\epsilon_j
     {\label{eq:mpc}}{}
    \addConstraint{ x_{i+1} } { = A x_i + B u_i }
    \addConstraint{ 0 }{ \leq v_i + \epsilon_0 }
    \addConstraint{ v_i - \epsilon_1 }{ \leq v_\mathrm{max} }
    \addConstraint{ u_\mathrm{min} \leq u_i }{ \leq u_\mathrm{max} }
    \addConstraint{\mathbb{P}\left(  s^\mathrm{pv}_{i} - s_i - T v_i - \epsilon_2 \geq d_\mathrm{min} \right) \geq \alpha_i}
    \addConstraint{ \epsilon_j }{ \geq 0, }
    \addConstraint{ \text{for} \ i=0,\dots,N-1 ; \ j=0,\dots,2 } 
\end{mini}
The control model follows a linear state-space representation. The state vector $x = [s, v, a]^\intercal$ contains the longitudinal position, velocity, and acceleration of the vehicle, respectively. The continuous-time dynamics are
\begin{equation}
\dot x = [v, a, (u-a)/\tau]^\intercal
\end{equation}
where $\tau$ is a first-order lag constant on the acceleration response of the vehicle following a control input. This model is then discretized via a zero-order hold with discretization step $\Delta t$. 
Bounded constraints exist on the admissible values of the control that can be selected: $u_\mathrm{min}$ is the minimal allowed acceleration, and $u_\mathrm{max}$ is the maximal allowed acceleration, as well as on the admissible values of the states: $v_\mathrm{max}$ is the maximal allowed velocity and $d_\mathrm{min}$ is the minimum allowed standstill gap.

\newcommand{\sv}{\mathrm{pv}}
The potential future actions of the PV are unknown and must be reasoned about. As in \cite{ard2021energy}, we probabilistically treat its motion as a stochastic process where the future actions that the PV can take follow from a Gaussian distribution $u_\sv \sim \mathcal N\left( \mu_\sv , \ \sigma_\sv\sigma_\sv^\intercal \right)$. 
We then constrain the minimum allowed longitudinal gap from the PV $d_\mathrm{front} := s^\mathrm{pv} - s$, 
where the admissible probability of satisfying the constraint at each stage must be greater than $\alpha_i \in [0.50, \ 1.0)$. 
Starting from an initial confidence $\alpha_0 = 0.\overline{99}$, $\alpha_i$ is loosened as the horizon progresses - so that conservativeness in the constraint about the more distant (and more uncertain) future is reduced. A discounting factor $\lambda = 0.95$ is introduced, such that $\alpha_i = \lambda^i \alpha_0 \ \forall \ i \in [1, N]$.

Using this optimal control structure, we propose to investigate the following three variations in this study. These variations arise from how $d_\mathrm{front}$ is evaluated through the prediction horizon and interpreted by the controller. Let $ \{s^\mathrm{pv}_{i}, v^\mathrm{pv}_{i}, a^\mathrm{pv}_{i}\}$ denote the reference PV trajectory fed to the ego MPC for the car following task. In the baseline controller (PCF), this trajectory is computed by assuming that only the current status (i.e., the states) of the preceding vehicle are available to the ego vehicle via the V2X channel. The reference position and velocity trajectory of the PV are then generated for the ego MPC using kinematics as in Eq. \eqref{eq:pcf}. 
\begin{equation} \label{eq:pcf}
\begin{aligned}
    a^\mathrm{pv}_{i} &=
    \begin{cases}
         a^\mathrm{pv}_{0}, & \text{if } 0 < v^\mathrm{pv}_{i} < v_{max} \\
        0, & \text{otherwise}
    \end{cases} \\
     v^\mathrm{pv}_{i+1} &= \min \left(\max\left(0, \ v^\mathrm{pv}_{i} + a^\mathrm{pv}_{i} \Delta t \right), \ v_\mathrm{max} \right), \\
     s^\mathrm{pv}_{i+1} &=  s^\mathrm{pv}_{i} +  v^\mathrm{pv}_{i} \Delta t + \frac{1}{2} a^\mathrm{pv}_{i} \Delta t^2.
\end{aligned}
\end{equation}
To make the indexing explicit for the remaining controllers, let $p$ denote the preview index of the transmitted PV trajectory at the ego controller. In the nominal intention-sharing case, the received preview is time-aligned with the ego prediction horizon, so that $p=i$. Under delayed reception, the preview index is shifted by $\gamma$ steps, so that $p=i+\gamma$.

\subsection{Threat Model for Denial-of-Service} \label{sec:dos_model}

We consider adversaries that degrade the wireless backhaul of connected autonomous driving systems by overwhelming the communications networking stack.
Unlike physical-layer jamming that immediately disrupts radio access, the objective is to induce persistent delay and jitter via resource exhaustion, forcing the cooperative controller to operate on stale information.
This scenario reflects safety-critical applications that fail gradually via timing violations rather than by hard disconnects. 

\begin{figure}
    \centering
    \includegraphics[width=\linewidth]{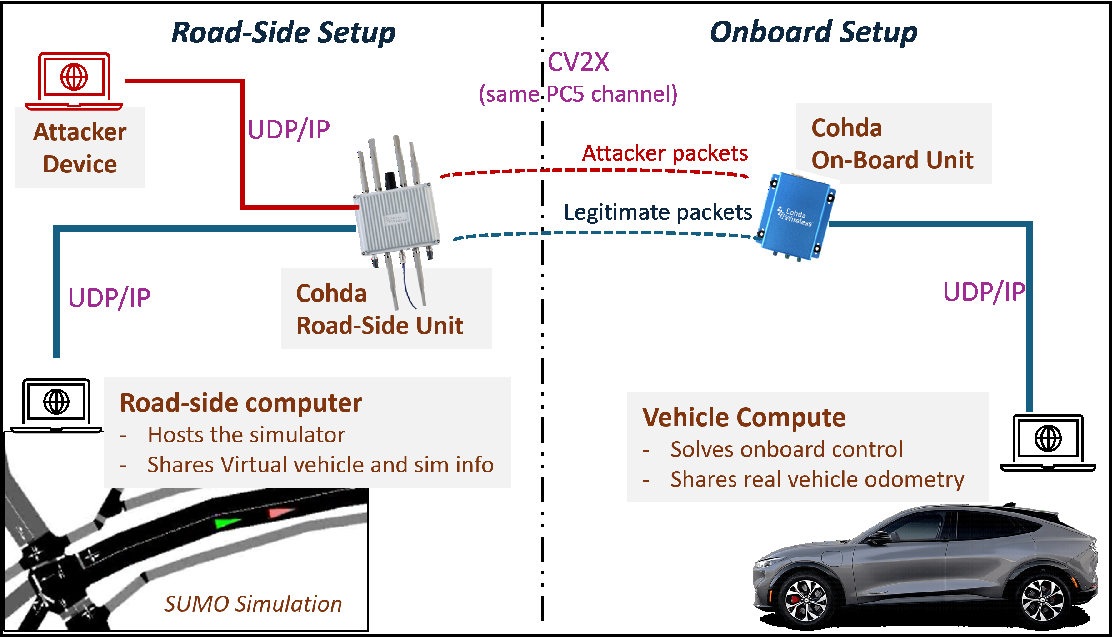}

    \caption{Vehicle in the loop experiment setup with a real CAV.}
    \label{fig:vil}
\end{figure}

\textit{System topology and attack location:}
Because the experimental platform is a simulation with vehicle in the the loop style setup, it uses an infrastructure-mediated communication path rather than direct V2V communication between two physical vehicles. As illustrated in Fig.~\ref{fig:vil}, the roadside computer hosts the SUMO simulator and the virtual vehicles. The status or intention message is transmitted over UDP/IP from the roadside computer to the Road-Side Unit (RSU), forwarded by the RSU over the C-V2X PC5 sidelink to the On-Board Unit (OBU), and then delivered to the ego-vehicle computer.
Ego-vehicle localization and status information follow the reverse path. The resulting information flow is therefore functionally V2I2V, with the RSU acting as the shared intermediary, and direct communication between two physical vehicles is not considered.

For this study, the adversary choice is inspired by the UDP based Denial-of-service attack method studied in \cite{tine2025real}. The adversary is located on an IP network from which the RSU-facing network services are reachable and assumed to be compromised.
Attack traffic is injected at the IP-facing interface of the RSU as surrounding (ghost) vehicle data. 
Thus, the networking stack of the RSU, and the associated processing and queuing resources are directly impacted, whereas the RSU--OBU PC5 information flow is affected indirectly through contention for shared channel resources.
The attacker sends malignant UDP messages with multiple connections and high transmission rates to the RSU interface to strain processing resources and drive queue buildup without triggering rate-based detection \cite{trkulja2020denialofserviceattackscv2xnetworks, rfc4732}.

\textit{Adversary capabilities:}
We consider an adversary that has privileged access to the roadside network. The adversary can generate flooding traffic, create many concurrent connections, and control their establishment rate, duration, and payload transmission rate. The adversary cannot modify the vehicle controller, inject control commands, alter legitimate messages, or modify their timestamps.

\textit{Affected information flows and attack effect:}
Attack traffic competes with legitimate processing and communication resources at the RSU and thus affects the infrastructure-mediated vehicle communication path. The resulting resource contention produces time-varying end-to-end delay and jitter in authentic cooperative-driving messages while communication remains available. For some attack scenarios, sudden bursts are also observed.
The consequence is not necessarily packet loss, but rather increased end-to-end latency. Vehicle control messages may then arrive several hundred milliseconds to multiple seconds late, violating the discretization assumptions of the controller.
Importantly, the RSU remains responsive and vehicles continue to exchange data, but under stale or jittery information. This degradation can enlarge inter-vehicle gap errors, and elevate collision risk.
For the $k$th received message, the end-to-end delay and corresponding control-step offset are represented by
\begin{equation}
\begin{aligned}
    D_k &= t^{\mathrm{rx}}_k-t^{\mathrm{tx}}_k
         = D_{0,k}+D_{q,k}+D_{\mathrm{j},k}
\end{aligned}
\label{eq:delay_model}
\end{equation}
where $t^{\mathrm{tx}}_k$ and $t^{\mathrm{rx}}_k$ are the synchronized transmission and reception timestamps, $D_{0,k}$ is the nominal propagation and processing delay, $D_{q,k}$ is the time-varying queuing and resource-contention delay, and $D_{\mathrm{j},k}$ denotes short-term service-time variation. The attack primarily increases $D_{q,k}$ and the variability represented by $D_{\mathrm{j},k}$.

\textit{Scope and assumptions:}
PC5 jamming or flooding, and control-command injection are outside the scope of this study. We consider delay-dominant attacks in which authentic, timestamped messages continue to arrive, although some may be delayed or lost. Accordingly, we evaluate whether intention sharing reduces control degradation relative to status-only sharing, rather than providing attack detection or guaranteed safety.

\begin{figure}
    \centering
    \includegraphics[width=\linewidth]{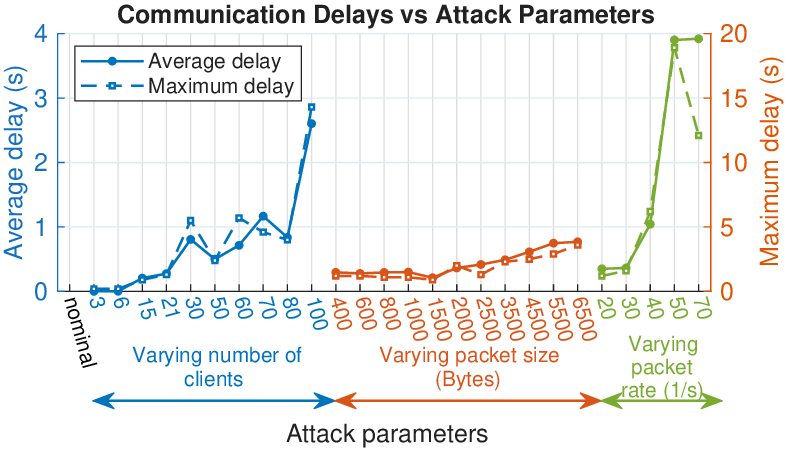}
    \caption{Sensitivity of the measured communication delay to attack
        parameters. The number of attacking clients, packet payload size, and
        per-client packet transmission rate are varied independently while the
        remaining parameters are held fixed. Filled solid curves show average delay (left Y axis),
        and open dashed curves show maximum delay (right Y axis).}
    \label{fig:sweeps_delay}
\end{figure}

To emulate realistic attack conditions, we first characterized the hybrid attacks on the Cohda MK6 hardware to produce round-trip delays. These measured delay profiles were subsequently used to calibrate the controlled delay-injection trials, ensuring the results are grounded in real-world network layer behavior while maintaining scenario repeatability. 
These were observed in the range of \unit[100]{ms}--\unit[2000]{ms} for reasonably severe attacks. 
Fig.~\ref{fig:sweeps_delay} shows the relationship between the attack parameters and the resulting communication delay when the number of attacking clients, packet payload size, and packet transmission rate are varied independently. The offered attack traffic is approximately proportional to $R_{\mathrm{off}}=N_c \cdot f_p \cdot L_p$, where $N_c$ is the number of attacking
clients, $f_p$ is the per-client packet transmission rate, and $L_p$ is the packet payload size in bytes. 
Therefore, larger packets increase the offered byte rate and, as the communication and processing resources become saturated, can increase queue occupancy, serialization delay, and packet loss. 
The packet-size sweep shows comparatively modest changes in delay over the smaller payload sizes, with a more noticeable increase only at the largest payloads tested. In contrast, increasing the number of attacking clients or the packet transmission rate produces substantially larger increases in both average and maximum delay. 
Packet-loss probability was not independently quantified in this experiment, so the present conclusions are limited to the measured end-to-end delay within the tested parameter ranges.


\subsection{Mitigation via Intention Sharing}

In the Predictive Car Following with Intention-sharing (PCF-I) formulation, the preceding vehicle periodically broadcasts not only its instantaneous states but also its predicted motion plan over the next $N$ stages of the MPC horizon. Because each vehicle is governed by a predictive controller, its optimized sequence of future accelerations, velocities, and positions constitutes an explicit representation of its short-term driving intention. Upon reception, these values replace the extrapolated kinematic estimates in the baseline PCF and serve as direct references in the follower's prediction model.

In the nominal case, where the preview is received on time, the ego controller uses $p=i$ and directly follows the transmitted trajectory:
\begin{equation} \label{eq:pcf-i}
\begin{aligned}
    a^\mathrm{pv}_{i} &=
    \begin{cases}
         a^\mathrm{pv}_{p}, & \text{if } 0 < v^\mathrm{pv}_{p} < v_{max} \\
        0, & \text{otherwise}
    \end{cases}\\
     v^\mathrm{pv}_{i}  &= v^\mathrm{pv}_{p}\\
     s^\mathrm{pv}_{i}  &= s^\mathrm{pv}_{p}
\end{aligned}
\end{equation}
where $\{s^\mathrm{pv}_{p}, v^\mathrm{pv}_{p}, a^\mathrm{pv}_{p}\}$ denotes the $p$-th preview state of the PV and $\{s^\mathrm{pv}_{i}, v^\mathrm{pv}_{i}, a^\mathrm{pv}_{i}\}$ denotes the reference trajectory used by the ego MPC. 
The follower directly tracks the acceleration, velocity, and position samples transmitted by the preceding vehicle. If the leader's velocity remains within allowable limits, its predicted acceleration is copied as the reference; otherwise, acceleration is clamped to zero to enforce kinematic bounds.

Compared with status-only sharing, this mechanism provides the ego controller with temporal context: the expected evolution of the leader's motion. Such preview information reduces dependence on perfect synchronization of messages because several future reference samples are already available locally. Hence, even when communication delays occur within a few discretization steps, the ego MPC can continue solving its optimization problem using the received trajectory segment without extrapolating outdated data. Intention sharing therefore acts as a form of time-domain redundancy that converts communication freshness requirements into prediction-accuracy requirements.
\begin{equation} \label{eq:pcf-ida}
\begin{aligned}
    p &= i + \gamma, \\
    \gamma &= \mathrm{Int}\!\left(\left|\frac{t_{\text{vehicle-clock}} - t_{\text{sim-clock}}}{\Delta t}\right|\right), \\
    a^\mathrm{pv}_{i} &=
    \begin{cases}
         a^\mathrm{pv}_{p}, & \text{if } p < N \ \text{and}\ 0 < v^\mathrm{pv}_{p} < v_\mathrm{max}, \\
        0, & \text{otherwise},
    \end{cases} \\
    v^\mathrm{pv}_{i} &=
    \begin{cases}
         v^\mathrm{pv}_{p}, & \text{if } p < N, \\
        0, & \text{otherwise},
    \end{cases} \\
    s^\mathrm{pv}_{i} &=
    \begin{cases}
         s^\mathrm{pv}_{p}, & \text{if } p < N, \\
         s^\mathrm{pv}_{N-1}, & \text{otherwise}.
    \end{cases}
\end{aligned}
\end{equation}
To further study the resilience, the PCF with Intention-sharing and Delay Awareness controller (PCF-IDA) augments PCF-I with explicit delay detection using timestamps embedded in V2X packets. If the received trajectory is delayed by $\gamma$ sampling steps, the ego controller aligns the preview by setting $p=i+\gamma$ as in Eq. \eqref{eq:pcf-ida}. When $\gamma>0$, the controller rolls the most recently received intention sequence forward by $\gamma$ stages, effectively compensating for latency. 
This ensures continuity of reference information even under transient communication stalls and biases the optimization toward maintaining larger inter-vehicle gaps when latency becomes severe.
Broadcast intentions represent the PV's planned motion at transmission and may become inaccurate after replanning or disturbances. Its effectiveness will eventually depend on the remaining preview being sufficiently accurate.

In this study, each vehicle transmits its planned trajectory at the same rate as the control update, i.e., \unit[20]{Hz}. 
Thus, a new intention message is broadcast every \unit[0.05]{s}. 
Each transmitted message contains $N=32$ predicted trajectory points compressed into a compact structure with timestamps for temporal alignment. 
The \unit[0.05]{s} broadcast period should be distinguished from the prediction-grid interval used to sample the preview trajectory. 
Following the MPC setup in \cite{ard2021energy}, the transmitted preview trajectory is sampled at \unit[0.5]{s} intervals; therefore, the $N=32$ preview points correspond to a \unit[16]{s} look-ahead.
This preview duration is sufficient to cover the largest induced delays considered in the experiments. 
Because the planned trajectory is already generated internally, intention sharing adds only fixed-size parsing and alignment before the unchanged Gurobi MPC. In the tested implementation, the solve time is under approximately \unit[25]{ms} on a \unit[4.8]{GHz} Intel Xeon, while \unit[272]{byte} messages at \unit[20]{Hz} required only \unit[43.52]{kbit/s}, indicating negligible added overhead.

Intention sharing differs fundamentally from conventional cooperative strategies that require consensus or negotiation among vehicles. It requires no iterative communication and is fully feed-forward, making it compatible with standard broadcast messaging. By embedding predictive information directly into transmitted packets, the follower's control law becomes less sensitive to occasional message delay. The next section demonstrates, through vehicle-in-the-loop experiments, how this predictive layer translates into measurable improvements in safety and stability under adversarial delay.

\begin{algorithm}[]
\caption{Mitigation with Intention Sharing (PCF–IDA)}
\label{alg:pcf-ida}
\begin{algorithmic}[1]
\STATE \textbf{Input:} Received intention sequence 
$\{s^{pv}_p, v^{pv}_p, a^{pv}_p\}_{p=0}^{N-1}$, timestamps $t_{\text{vehicle}}$, $t_{\text{sim}}$, discretization step $\Delta t$
\STATE \textbf{Compute delay offset:} 
$\gamma \gets \text{Int}\!\left(|(t_{\text{vehicle}} - t_{\text{sim}})/\Delta t|\right)$
\FOR{$i = 0$ to $N-1$}
    \IF{$i+\gamma < N$}  
        \STATE \COMMENT{roll forward with $\gamma$}
        \STATE $v^{pv}_i, s^{pv}_i \gets v^{pv}_{p+\gamma}, s^{pv}_{p+\gamma}$
        \STATE $a^{pv}_i \gets a^{pv}_{p+\gamma} \text{ or } 0$ 
    \ELSE 
    \STATE \COMMENT{Pad with stop assumption}
        \STATE $s^{pv}_i \gets s^{pv}_{N-\gamma}$  
    \ENDIF
\ENDFOR
\STATE Solve ego MPC Eq. \eqref{eq:mpc} with $\{s^{pv}_i,v^{pv}_i,a^{pv}_i\}$ as preview references.
\STATE \textbf{Output:} Delay-compensated preview trajectory
\end{algorithmic}
\end{algorithm}

\section{Experimental Setup}
The resilience of intention-sharing is evaluated using a vehicle-in-the-loop testbed as depicted in Fig. \ref{fig:vil}. The physical platform is a drive-by-wire electric vehicle equipped with RTK GPS, IMU, and LiDAR, managed via a \texttt{ROS/ROS2} stack. A roadside computer hosts a digital twin of the vehicle within a SUMO-based microscopic traffic simulator \cite{dlr127994} to provide a time synchronized, controlled environment.

\begin{figure}
    \centering
    \includegraphics[width=\linewidth]{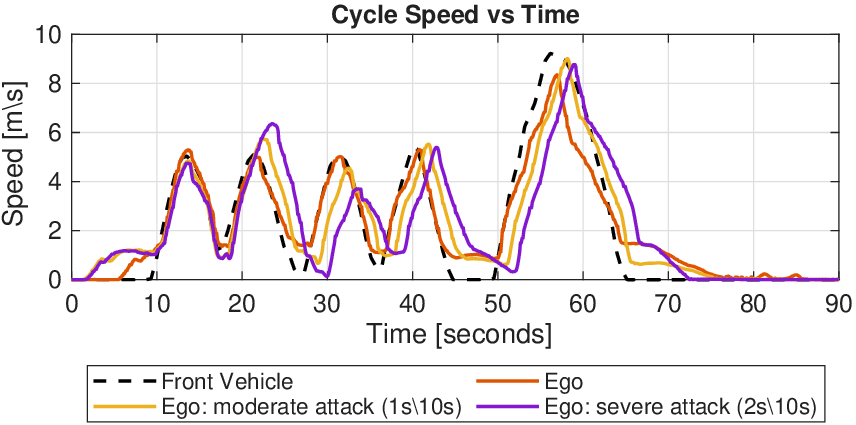}

    \caption{Speed trajectories of the real ego CAV with the PCF controller and the preceding virtual CAV follow the US06 reference driving cycle, under nominal and attack conditions.}
    \label{fig:cycle}
\end{figure}

The communication channel is enabled using a \texttt{Cohda MK6} CV2X/DSRC stack between the vehicle and the road-side infrastructure using the OBU and RSU devices, as explained in the topology description in Section \ref{sec:dos_model}.
The clocks of the simulator, RSU, OBU, and vehicle-processing devices are synchronized to GPS clock time. 
Longitudinal car-following scenarios with two configurations are considered: a two-vehicle car-following case and a three-vehicle car-following case, with the real CAV equipped with the physical V2X stack acting as the second vehicle in both configurations. The three-vehicle case extends the evaluation beyond the single-preceding-vehicle configuration and examines whether the observed controller trends persist when the real CAV is embedded within a longer longitudinal interaction. 
These configurations were selected to isolate the effects of the communication strategy and adversarial delay while holding the maneuver type, communication path, and physical test environment consistent across PCF, PCF-I, and PCF-IDA. 
This controlled evaluation enables direct and repeatable comparison of the three controllers using a real CAV and physical V2X communication layer.
The preceding simulated vehicle executes a scaled-down segment of the US06 driving cycle \cite{epaUS06} over a \unit[230]{m} test lane, as shown in Fig.~\ref{fig:cycle}.
The cycle includes multiple acceleration and deceleration events, allowing the controllers to be evaluated under delays introduced at different phases of the longitudinal maneuver. Real-time RTT measurements during the attacks confirmed communication delays exceeding \unit[2]{s}, which affect controller feedback and temporal alignment, as shown in Fig.~\ref{fig:cycle}.

\section{Results and Discussion} \label{sec:results}
This section evaluates the impact of adversarial delay on cooperative driving using the three controller variants in the two- and three-vehicle car-following configurations.

\subsubsection{Aggregate Impact}
\begin{figure}
    \centering
    \includegraphics[width=\linewidth]{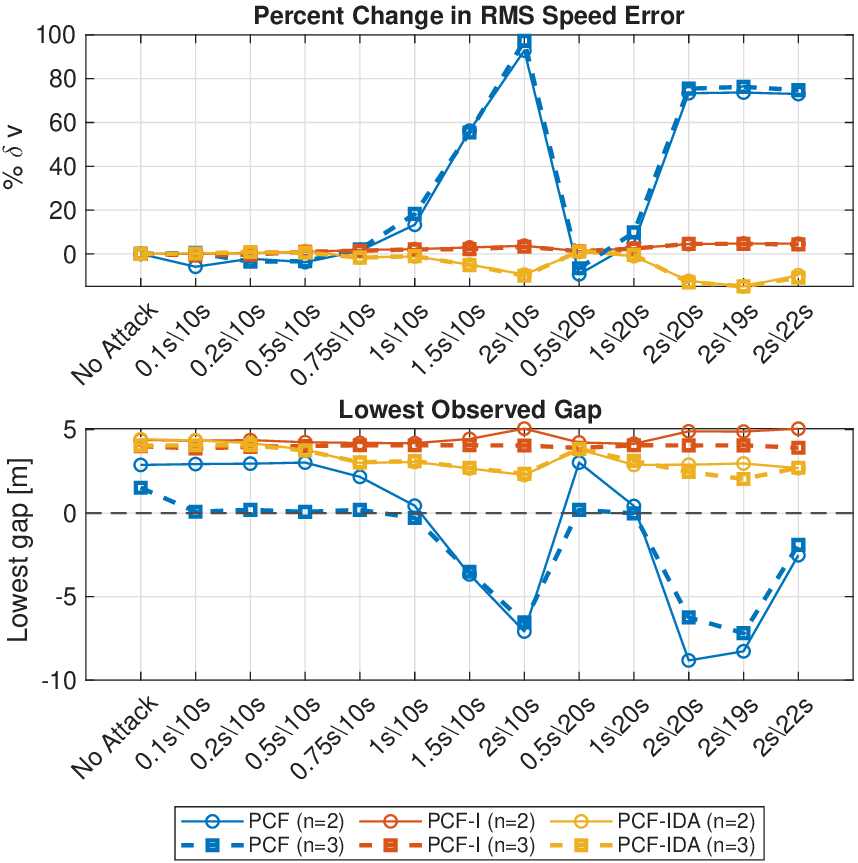}
    \caption{Impact of the delay-inducing attacks on controller performance across the tested scenarios. Each scenario is denoted by $d/t_a$, where $d$ is the effective communication delay and $t_a$ is the delay-onset time during the driving cycle; for example, ``2~s/22~s'' denotes a \unit[2]{s} delay introduced at $t=22$~s. The vertical axes show the percentage changes in RMS speed-tracking error and the lowest observed inter-vehicle gap relative to the corresponding no-attack condition. The solid line plots represent the performance in two-vehicle configuration and the dashed lines represent the three-vehicle configuration}
    \label{fig:impact}
\end{figure}

Fig. \ref{fig:impact} visualizes the overall impact severity of the delay inducing attack, using the normalized Root Mean Square (RMS) speed and lowest observed inter-vehicle distance gaps tested across $13$ attack scenarios for each of the chosen car-following configurations.
The ``No Attack'' scenario represents the corresponding nominal baseline. Each plot line corresponds to a controller variant, and the solid and dashed lines correspond to the two-vehicle and three-vehicle car-following configurations, respectively. 
In the first plot, the vertical-axis values represent the percentage change relative to that controller's corresponding no-attack condition. Accordingly, the reported ranges below denote the minimum and maximum percentage changes observed across the tested attack-delay scenarios. Negative percentages indicate that the error-change metric was slightly lower than in the corresponding no-attack trial.
For the three-vehicle configuration, the percentage RMS speed tracking error change shown in the first subplot is averaged over the two successive car-following interactions in the vehicle chain.

Under nominal (non-attack) conditions and for small delays under \unit[100]{ms}, all controllers perform as expected.
However, under attacks and delays of greater than \unit[100]{ms}, PCF with only status sharing exhibits significant degradation and speed tracking errors rise up to 100\%. This lag yields more aggressive and jerky control inputs and increases the overall control effort as well, when compared to nominal behavior.
The effect of communication delay is most pronounced for the baseline PCF controller. 
Across the tested scenarios, its RMS speed-tracking error changes by $-9.3\%$ to $92.9\%$, compared with $0.01\%$ to $4.8\%$ for PCF-I and $-14.8\%$ to $1.3\%$ for PCF-IDA.
Thus, both intention-sharing variants maintain speed-tracking behavior substantially closer to their respective nominal conditions as communication delay increases.
The corresponding RMS inter-vehicle-gap changes, not shown in Fig.~\ref{fig:impact}, range from $-44.4\%$ to $-2.9\%$ for PCF, from $-0.3\%$ to $2.7\%$ for PCF-I, and from $-21.9\%$ to $0.03\%$ for PCF-IDA. These trends are consistent with the minimum-gap results discussed in the following subsection.

Communication delay also increases actuation demand. Across the two- and three-vehicle cases, the RMS control-effort change ranges from $12.6\%$ to $101.6\%$ for PCF, from $2.5\%$ to $76.0\%$ for PCF-I, and from $9.0\%$ to $109.4\%$ for PCF-IDA. PCF-I generally limits this increase relative to baseline PCF, whereas PCF-IDA can require greater corrective effort because its delay-aware fallback favors conservative gap preservation.
The solid and dashed curves in Fig.~\ref{fig:impact} also show that introducing a second car-following interaction does not materially change the relative controller trends. In both configurations, status-only PCF exhibits the strongest degradation with increasing delay, while PCF-I and PCF-IDA remain substantially less sensitive with respect to their nominal cases. The three-vehicle case therefore provides additional evidence that the observed benefit is not restricted to a single follower interaction.

\begin{table*}[]
\centering
\caption{Qualitative summary of the main experimental observations under adversarial delay}
\label{tab:summary_results}
\begin{tabularx}{\textwidth}{
    >{\hsize=0.30\hsize}X
    >{\hsize=1.2\hsize}X
    >{\hsize=1.2\hsize}X
    >{\hsize=1.2\hsize}X
}
\hline
\textbf{Controller} & \textbf{Observed behavior under delay} & \textbf{Observed collision outcome} & \textbf{Interpretation} \\
\hline
PCF & Strong lag in tracking; degraded response to delayed PV information & Rear-end collisions occur under severe delay & Status-only sharing is highly sensitive to stale information \\
PCF-I & Smoother motion; better anticipation of leader behavior & No collisions in tested scenarios & Intention preview provides useful temporal redundancy \\
PCF-IDA & Conservative response; prioritizes gap preservation over tracking aggressiveness & No collisions in tested scenarios & Timestamp alignment and fallback logic improve robustness \\
\hline
\end{tabularx}
\end{table*}

In summary, PCF-I leverages predictive context to minimize deviations and mitigate braking variations. PCF-IDA rolls the delayed previews forward to the worst case scenario in terms of time to collision, sustaining conservative gaps under the tested severe latency conditions.
Fig. \ref{fig:cycle} illustrates the longitudinal trajectories where baseline PCF suffers from overshoots during acceleration peaks and braking events under delay. 
PCF-I maintains tracking with comparatively moderate changes in control effort, whereas PCF-IDA prioritizes conservative gap preservation and may consequently require greater corrective control effort under severe delay.

\subsubsection{Vehicle Gap and Collision Analysis}
Gap maintenance is evaluated via time-series distance gaps and collision incidents across the multiple trials per controller-delay combination. 
We flag collision when the measured gap violates the effective minimum $\big\{\,d_\mathrm{front} < d_{\min}^{\mathrm{bumper}}\,\big\}$.
In the event of collisions in the experiments, the vehicles were allowed to continue on after the bumper impact to expose all potential collision risks and events through the trials. 
The second plot in Fig. \ref{fig:impact} shows the lowest observed gap among all vehicle interactions in the given scenario. The baseline PCF controller experiences frequent rear-end collisions at delays exceeding the threshold of around \unit[1]{sec}, and seven trials in each configuration resulted in crashes. This controller fails to accelerate or brake in sync under attacks and yields late braking and excessive lag. 
In contrast, the PCF-I preserves sufficient temporal context to avoid collisions while still maintaining reasonable tracking. 
The PCF-IDA yields collision-free operation across the conducted trials by adopting a worst-case fallback upon detecting timestamp mismatches. This conservative fallback variant maintains vehicle gaps, even under maximum tested delay conditions (\unit[2.0]{s}). Minimum following distances consistently remain close those of PCF-I. 
Since these numbers depend on the scenario type, and the time at which the attack is launched, our trials include attack launches at different times of $t = 10, 19, 20, \text{ and } 22~\mathrm{s}$ of the cycle.

This trend suggests that intention sharing not only supports performance but can serve as a practical resilience layer in connected vehicle systems under non-ideal communication conditions.

These findings are limited to the tested single-preceding-vehicle car-following scenarios and do not constitute a theoretical safety guarantee. Collision risk may remain under communication loss, delays beyond the preview horizon, inaccurate intentions, unexpected emergency maneuvers, or unmodeled disturbances. Evaluation of such conditions, as well as lane changes, intersections, and multi-vehicle interactions, is left for future work.

\section{Conclusion}
This study provides empirical evidence that intention sharing communication strategy enhances the resilience of connected autonomous driving under DoS-induced communication delays. The conducted Vehicle-in-the-loop (VIL) experiments demonstrate that while baseline status-sharing controllers suffer from degraded tracking and high collision rates, the intention-sharing variants avoid collisions across the 13 scenarios for each of the vehicle configurations. These results indicate that status sharing alone can be insufficient for robust cooperative driving in adversarial environments. By converting message freshness requirements into a prediction accuracy problem, intention-sharing architectures provide a practical foundation for safeguarding connected vehicle ecosystems against network-layer degradation.
\bibliographystyle{IEEEtran}
\bibliography{VT-2026-03085_R1_final_refs}

\end{document}